\documentclass[preprint,12pt]{elsarticle}
\usepackage{amsmath,amssymb}
\usepackage{graphicx}
\usepackage{booktabs}
\usepackage{siunitx}
\usepackage{hyperref}
\journal{Nature Communications}

\begin{document}

\begin{frontmatter}

\title{Indirect monitoring of fast-charge cycling behaviour of an energy-storage device---analysis of ambient temperature variations}

\author[1]{Pertti O. Tikkanen\corref{cor1}\fnref{orcid}}
\ead{pertti.tikkanen@gmail.com}
\cortext[cor1]{Corresponding author}
\fntext[orcid]{ORCID: 0000-0003-1422-3320}
\address[1]{Department of Physics, University of Helsinki, Finland}

\begin{abstract}

I present a reanalysis of temperature data from a 
publicly available certified laboratory report that documented the self-discharging behaviour of
an energy-storage device during 10 days. Graphs of temperature variations of both the tested device
itself and the test chamber (fume hood) were given
mainly for monitoring without further analysis, and variations in the 
ambient temperature signal were attributed to "other cells being cycled simultaneously in the same fume hood".
I show that the ambient temperature signal alone---together with some quite mild and reasonable
assumptions---allow to extract previously unpublished information on the simultaneously run test on the other cells:
1) the number of charge/discharge cycles 2) the cycle period, 3) the charge/discharge half-cycle
asymmetry, and---most significantly---evidence that 4) the mentioned "other devices" completed 338 full
charge/discharge cycles at 3C rate at room temperature without any detectable thermal degradation signature.

Five analysis segments are extracted from the single \SI{254}{\hour} record,
together covering \SI{\approx 330} cycles.
All segments exhibit the same thermal oscillation: a \SI{\sim 40}{\minute}
full cycle composed of a \SI{\sim 22}{\minute} charge half-cycle and a
\SI{\sim 18}{\minute} discharge half-cycle.
The \SI{\sim 4}{\minute} difference $\Delta = T_{\mathrm{charge}} -
T_{\mathrm{discharge}}$, stable throughout the run, may point to an intentional
relaxation pause in the cycling protocol.
The Fourier spectrum is a clean harmonic series fully explained by this
alternating-period structure.
Amplitude and period are stable over \SI{254}{\hour} with no sign of device
degradation detectable from this indirect measurement alone.
\end{abstract}


\end{frontmatter}

%\linenumbers

%% ─────────────────────────────────────────────────────────────────────────────
\section{Introduction}
\label{sec:intro}

Thermal characterisation of electric energy-storage devices during cycling
is central to understanding degradation, validating thermal management,
and ensuring safe operation~\cite{Waldmann2014,Bandhauer2011}.
The usual approach---use of thermocouples, IR cameras, or
calorimeters---yield accurate data but require dedicated
instrumentation and often physical access to the device.

Here I show that a fume-hood air-temperature sensor can resolve
individual charge/discharge half-cycles and track the thermal amplitude over
hundreds of cycles without any modification to the device or the test set-up.
Although the fume-hood air acts as a dampening low-pass thermal filter,
the essential features of the saw-tooth-wave temperature oscillations remain intact, allowing
standard signal processing.

The current study doesn't report any new experiments or new experimental data, the
presented results are based solely on a detailed reanalysis of temperature graphs shown
in Figures 2, 3, and 4 of the recently published report.\cite{originalreport} 
The primary purpose of that certified laboratory test report
by VTT Technical Research Centre of Finland, commissioned by Donut Lab, was to 
characterise the self-discharge performance of the Donut Lab Solid-State 
Battery V1. This is the third in a series of VTT certified test reports on 
Donut Lab solid-state cells.\cite{vttreport1,vttreport2,originalreport}
The experimental details of the test 
setup, instrumentation and measurement protocol are described in full in 
the report~\cite{originalreport} and are not repeated here.

The self-discharge test protocol consisted of monitoring the voltage drop of 
cell DL1 charged to 50\% state of charge (SOC) in 10-second intervals over a 
10-day period, and a standard discharge measurement at the end confirmed
that the cell retained  97.7\% of its original SOC. This demonstrated that the 
device exhibits battery-like behaviour rather than super-capacitor-like 
behaviour. As a part of the standard test protocol, a temperature sensor was
directly attached on the device and gave
additional information on short-term and long-term thermal stability of the device but had
minor importance to the primary self-discharge result. The single remark in the
report---\textit{variations in ambient temperature
were caused by other cells being cycled in the same fume hood}---and the clearly
observable oscillations in the ambient temperature graph indicated that a careful
reanalysis might reveal some information on the behaviour of the other devices.

Written permission to use the data was granted by both VTT and Donut Lab. No details of
any of the tested energy-storage devices were available, I report only what
the ambient temperature data of the fume hood reveals.

%% ─────────────────────────────────────────────────────────────────────────────
\section{Data and signal processing}
\label{sec:methods}

\subsection{Data extraction and segment definition}
\label{sec:data}

Temperature traces were extracted from Figures 2, 3, and 4 in
ref.~\cite{originalreport} in two steps: 1) by converting manually the graphs
of the original pdf file into SVG format using Inkscape~\cite{Inkscape}, and 2) by parsing the embedded path data
with the help of AI (Anthropic Claude v1.1.6452, Sonnet 4.6).
The graphs were resampled using a sampling interval  \SI{\approx 16}{\second}
($f_s \approx \SI{3.75}{\milli\hertz}$), because of the varying node spacing of the SVG paths didn't allow
ordinary Fourier analysis.
The good quality of the extraction process is demonstrated in Figures \ref{fig:rawA} and \ref{fig:rawB}, the latter reproducing the
details in the original Figure 2 of  ref.~\cite{originalreport}.
All three original figures contain also the record (red curve) of an temperature
sensor at the self-discharge test cell, showing rather large transient peaks at the
record boundaries consistent with a 1C
charge-recharge of that cell.
Both temperature tracks share the same long-term drift. The higher average level of the fume-hood temperature (blue curve) reflects
the mean heating power from the cycling of the other cells, balanced with thermal loss to the environment. The two approximately
12-hour smooth regions (at $t \approx 21$~h and $t \approx 168$~h in the blue curve) in
Figure \ref{fig:rawB} indicate that the cycling was interrupted twice for about 12 hours.
% whose cause is unstated in the report.

Five analysis segments selected for the analysis are defined in  Table~\ref{tab:periods}:
two \emph{zoomed} windows taken directly from the high-resolution
figures---ZE ($t = \SIrange{0}{20.5}{\hour}$, $N = 4{,}679$) and
ZL ($t = \SIrange{234.5}{254}{\hour}$, $N = 4{,}459$), separated by
\SI{\approx 214}{\hour}---and three
\emph{overview} segments extracted from the full-record figure: OE
($t = \SIrange{0}{21}{\hour}$), OI ($t = \SIrange{36}{168}{\hour}$), and
OL ($t = \SIrange{185}{254}{\hour}$).

\subsection{Drift removal}
\label{sec:detrend}

The slow baseline drift
%(presumably diurnal room-temperature variation and fume-hood equilibration)
was removed by fitting a clamped cubic spline---zero
first-derivative boundary conditions, \SI{1}{\hour} moving-average
pre-smoothing, \SI{30}{\minute} subsampling---to each segment independently (except for
segment OI).
For segment OI, the contact sensor on the self-discharge cell was linearly scaled to match the
fume-hood channel in the two interruption windows (gap~1: $t =
\SIrange{22}{35}{\hour}$; gap~2: $t = \SIrange{169}{184}{\hour}$), where
cycling had ceased:
\begin{equation}
  T_{\mathrm{ref,scaled}} = 1.248\,T_{\mathrm{ref}} - \SI{5.73}{\celsius},
  \quad r = 0.71.
  \label{eq:scaling}
\end{equation}
This scaled reference was subtracted globally across OI; a residual clamped
spline removed the remaining slow variation.
The detrended signal $\Delta T(t)$ has $\sigma = \SIrange{0.52}{0.57}{\celsius}$
across all five segments, with residual slow-frequency power below
$0.1\,\%$ of the cycling-band power in every case.

\subsection{Spectral and peak analysis}

The one-sided DFT amplitude spectrum of each detrended segment was computed
on the uniformly sampled record.
Peaks and troughs of $\Delta T(t)$ were located using a prominence threshold
of \SI{0.10}{\celsius} and a minimum inter-peak spacing of \SI{12}{\minute};
consecutive peak-to-peak intervals were partitioned into even- and odd-indexed
sub-sequences to extract the alternating half-cycle durations.

%% ─────────────────────────────────────────────────────────────────────────────
\section{Results}
\label{sec:results}

\subsection{Raw signal and drift baselines}

\begin{figure}[htbp]
  \centering
  \includegraphics[width=\columnwidth]{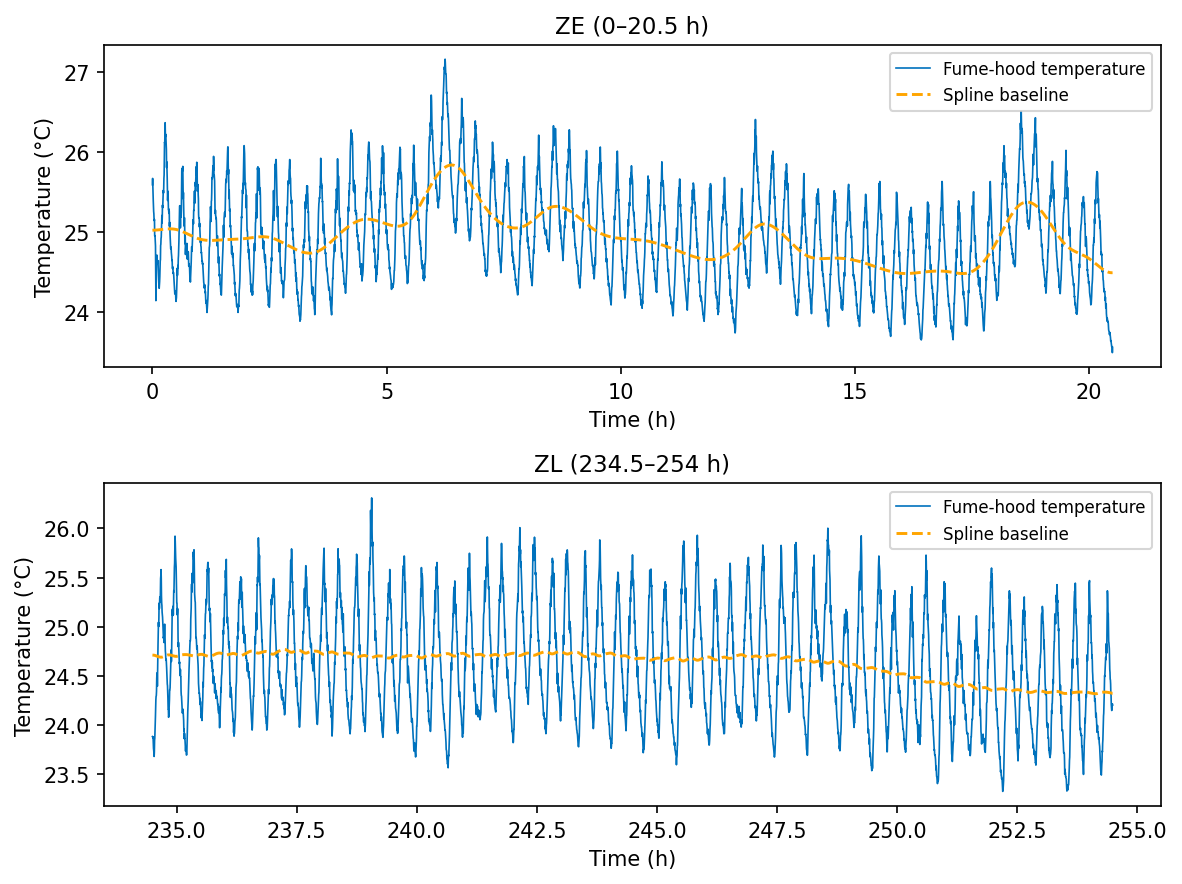}
  \caption{Zoomed windows ZE (top) and ZL (bottom): raw fume-hood temperature
           (blue) and cubic-spline drift baseline (orange dashed).}
  \label{fig:rawA}
\end{figure}

\begin{figure}[htbp]
  \centering
  \includegraphics[width=\columnwidth]{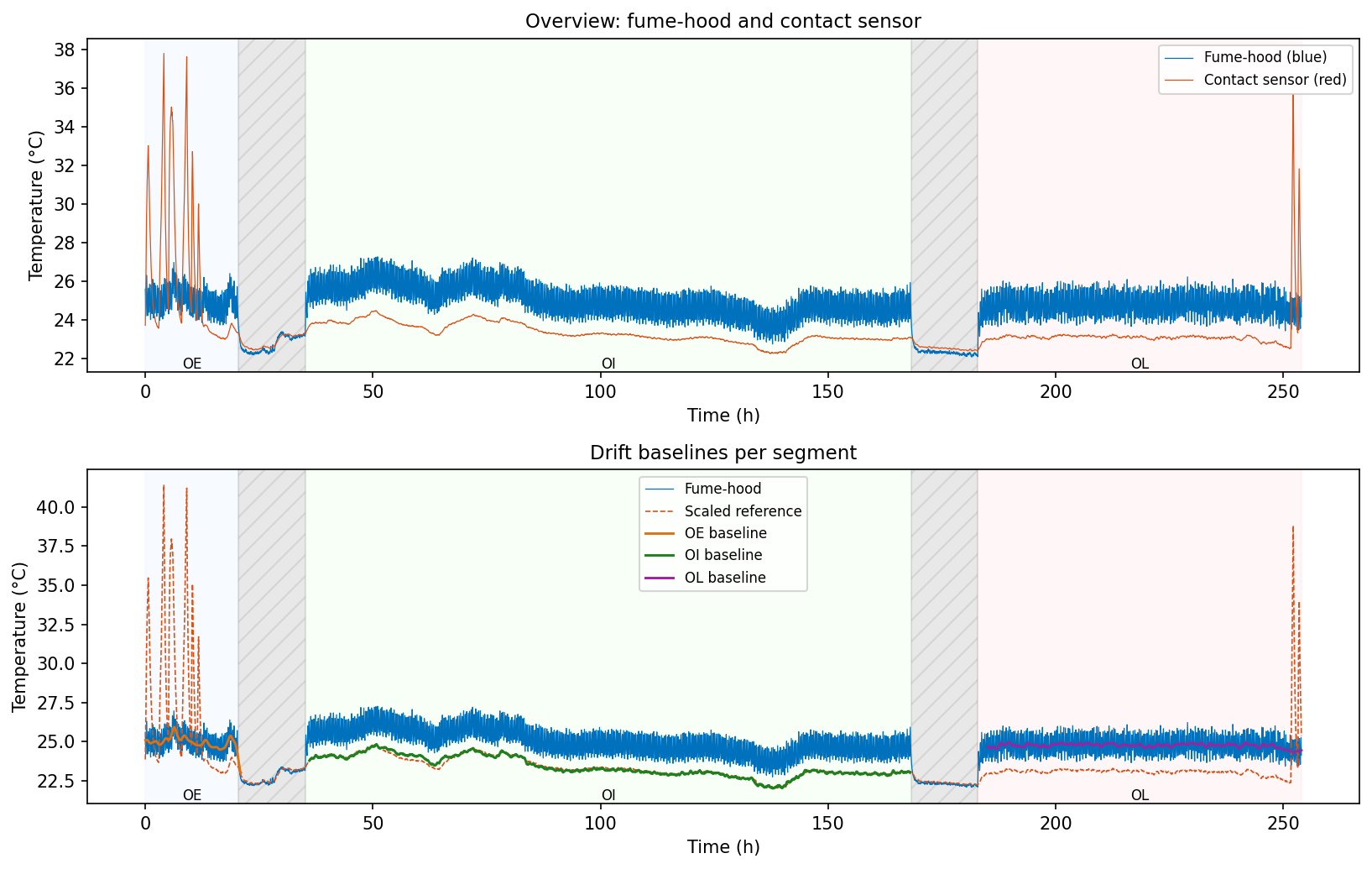}
  \caption{Full \SI{254}{\hour} record. Top: fume-hood temperature (blue) and
           contact sensor on the self-discharge cell (red); shaded bands are
           analysis segments OE, OI, OL; hatched regions are gap windows.
           Bottom: blue signal with scaled reference (red dashed) and fitted
           drift baselines (coloured lines) per segment.}
  \label{fig:rawB}
\end{figure}

The cycling oscillation (\SI{\sim 1}{\celsius} amplitude) is visible in all
five segments riding on a slowly varying background (Figs.~\ref{fig:rawA}
and~\ref{fig:rawB}).
The spline baseline varies by \SI{1.4}{\celsius} over ZE and \SI{0.6}{\celsius}
over ZL; the contact-sensor-aided baseline captures the larger excursions
around the interruptions in OI.

\subsection{Spectral fingerprint}

\begin{figure}[htbp]
  \centering
  \includegraphics[width=\columnwidth]{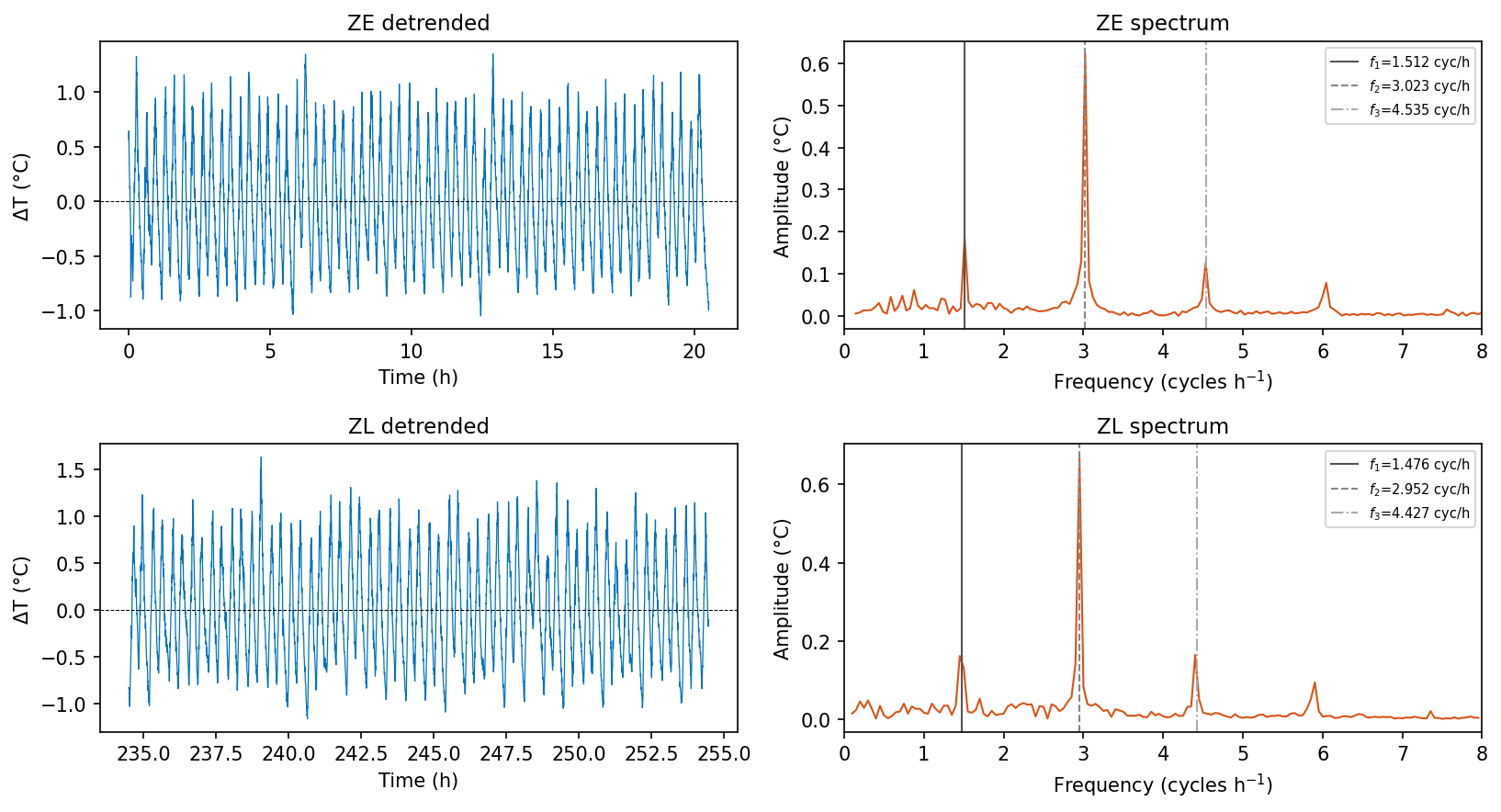}
  \caption{Detrended $\Delta T$ (left) and DFT amplitude spectra (right) for
           ZE (top) and ZL (bottom). Dashed lines mark the harmonic series
           $f_n = nf_1$.}
  \label{fig:fftA}
\end{figure}

\begin{figure}[htbp]
  \centering
  \includegraphics[width=\columnwidth]{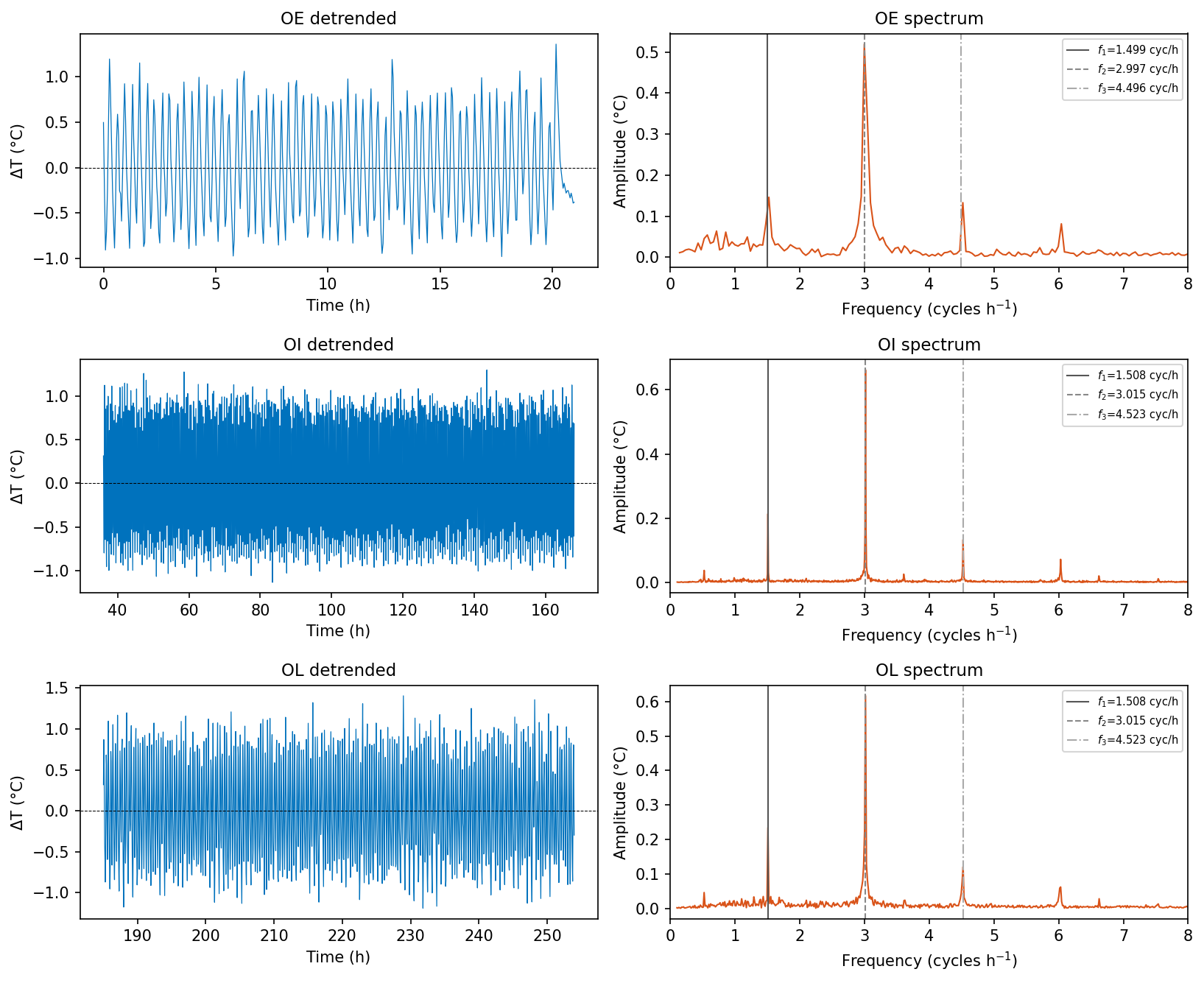}
  \caption{As Fig.~\ref{fig:fftA} for OE (top), OI (middle), OL (bottom).}
  \label{fig:fftB}
\end{figure}

After detrending, all five segments share the same spectral fingerprint
(Figs.~\ref{fig:fftA} and~\ref{fig:fftB}): a dominant peak at
$f_2 \approx \SI{3.0}{\per\hour}$ (period \SI{\approx 20}{\minute}, amplitude
\SIrange{0.52}{0.68}{\celsius}), a sub-dominant peak at $f_1 \approx
\SI{1.5}{\per\hour}$ (\SI{\approx 40}{\minute}), and a third at $f_3 \approx
\SI{4.5}{\per\hour}$ (\SI{\approx 13}{\minute}).
The integer ratios $f_1 : f_2 : f_3 = 1 : 2 : 3$ identify a harmonic series
with fundamental $T_{\mathrm{cycle}} \approx \SI{40}{\minute}$
(Section~\ref{sec:harmonics}).

\subsection{Alternating half-cycle structure and amplitude stability}

\begin{figure}[htbp]
  \centering
  \includegraphics[width=\columnwidth]{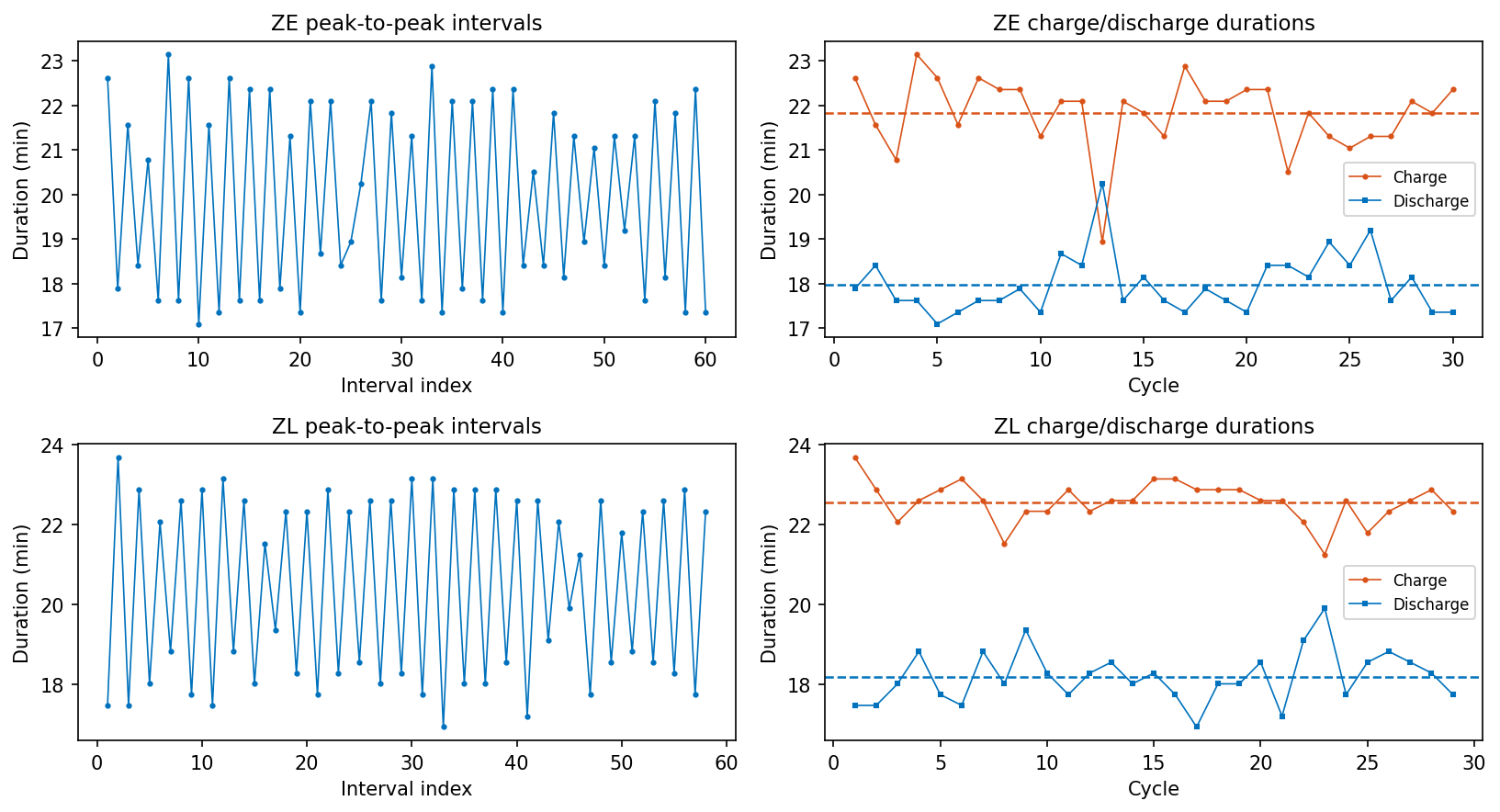}
  \caption{Peak-to-peak interval sequence (left) and per-cycle charge /
           discharge durations (right) for ZE and ZL.}
  \label{fig:intervalsA}
\end{figure}

\begin{figure}[htbp]
  \centering
  \includegraphics[width=\columnwidth]{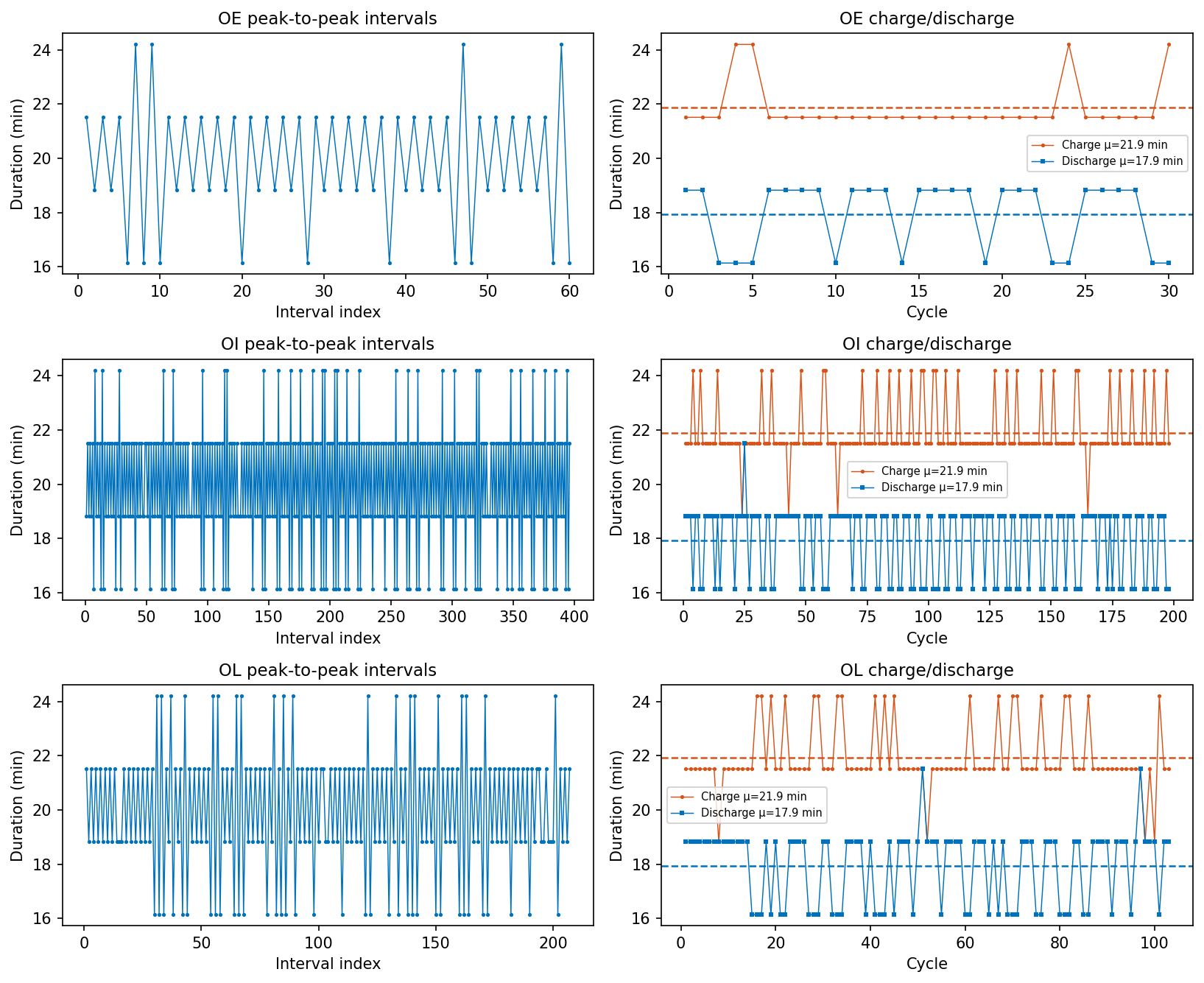}
  \caption{As Fig.~\ref{fig:intervalsA} for OE, OI, OL.}
  \label{fig:intervalsB}
\end{figure}

Because the variations in the temperature signal were known to be caused by other
cells being cycled in the same chamber, it was worth studying the alternating half-cycle structure in more detail.
Consecutive half-cycles alternate between a longer charge phase
$T_{\mathrm{charge}}$ and a shorter discharge phase $T_{\mathrm{discharge}}$
with small cycle-to-cycle scatter (Figs.~\ref{fig:intervalsA}
and~\ref{fig:intervalsB}, Table~\ref{tab:periods}).
The values $T_{\mathrm{charge}} \approx \SI{22}{\minute}$ and
$T_{\mathrm{discharge}} \approx \SI{18}{\minute}$ are consistent with 3C
cycling ($\SI{60}{\minute}/3 = \SI{20}{\minute}$ per half-cycle).
The difference $\Delta = T_{\mathrm{charge}} - T_{\mathrm{discharge}} \approx
\SI{4}{\minute}$ is reproducible to within \SI{0.5}{\minute} across all
five segments spanning the full \SI{254}{\hour} run. Note that the inferred rate 3C is rather conservative
lower limit, higher rates might also be consistent with the temperature data
if longer relaxation times were assumed.

Peak amplitudes exceed trough amplitudes by \SIrange{19}{31}{\percent} in all
segments (Table~\ref{tab:amplitude}), reflecting
the asymmetric heating/cooling kinetics of the charging-recharging cycle of the device.
Linear amplitude trends within each segment are negligible
($|\text{slope}| < \SI{2e-3}{\celsius\per\hour}$).

\begin{table}[htbp]
  \centering
  \caption{Cycle-period parameters with one-sigma uncertainties.
           $\Delta = T_{\mathrm{charge}} - T_{\mathrm{discharge}}$.
           Cycle counts: ZE and ZL $\approx\!30$ each;
           OE~30, OI~198, OL~104.}
  \label{tab:periods}
  \sisetup{separate-uncertainty=true}
  \begin{tabular}{l S[table-format=2.2(1)] S[table-format=2.2(1)]
                    S[table-format=2.2(1)] S[table-format=2.2(1)]
                    S[table-format=2.2(1)]}
    \toprule
    & \multicolumn{2}{c}{Zoomed} & \multicolumn{3}{c}{Overview} \\
    \cmidrule(lr){2-3}\cmidrule(lr){4-6}
    Parameter & {ZE} & {ZL} & {OE} & {OI} & {OL} \\
    \midrule
    $T_{\mathrm{charge}}$ (min)
      & 21.82(81) & 22.56(49) & 21.87(91) & 21.89(108) & 21.93(122) \\
    $T_{\mathrm{discharge}}$ (min)
      & 17.98(66) & 18.20(65) & 17.93(127) & 17.92(130) & 17.94(137) \\
    $T_{\mathrm{cycle}}$ (min)
      & 39.80 & 40.76 & 39.80 & 39.81 & 39.87 \\
    $\Delta$ (min)
      & 3.84 & 4.36 & 3.94 & 3.98 & 3.99 \\
    $T_{\mathrm{charge}}/T_{\mathrm{discharge}}$
      & 1.214 & 1.240 & 1.220 & 1.222 & 1.222 \\
    \bottomrule
  \end{tabular}
\end{table}

\begin{figure}[htbp]
  \centering
  \includegraphics[width=\columnwidth]{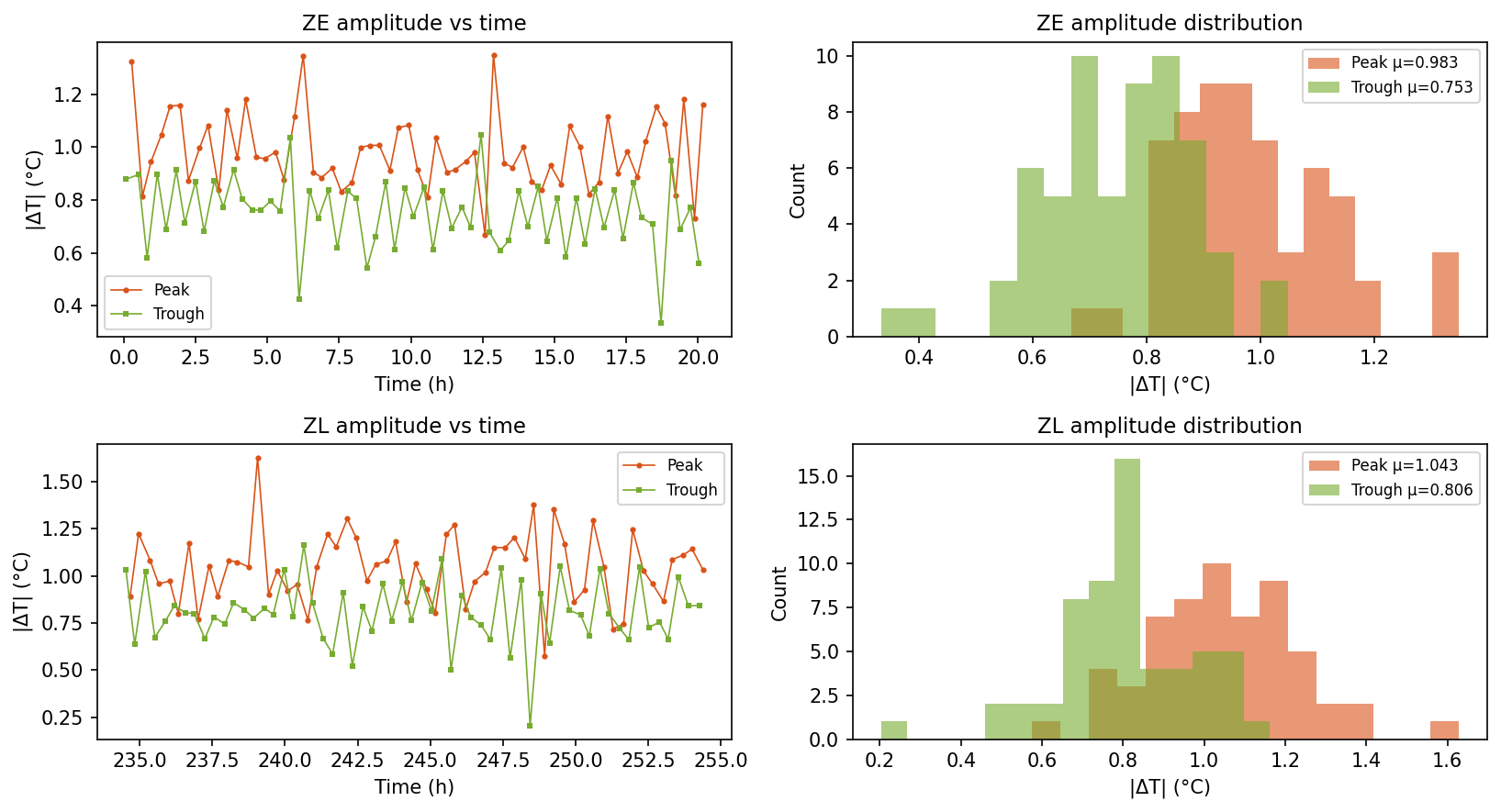}
  \caption{Peak (red) and trough (green) amplitudes and their distributions
           for ZE and ZL.}
  \label{fig:ampA}
\end{figure}

\begin{figure}[htbp]
  \centering
  \includegraphics[width=\columnwidth]{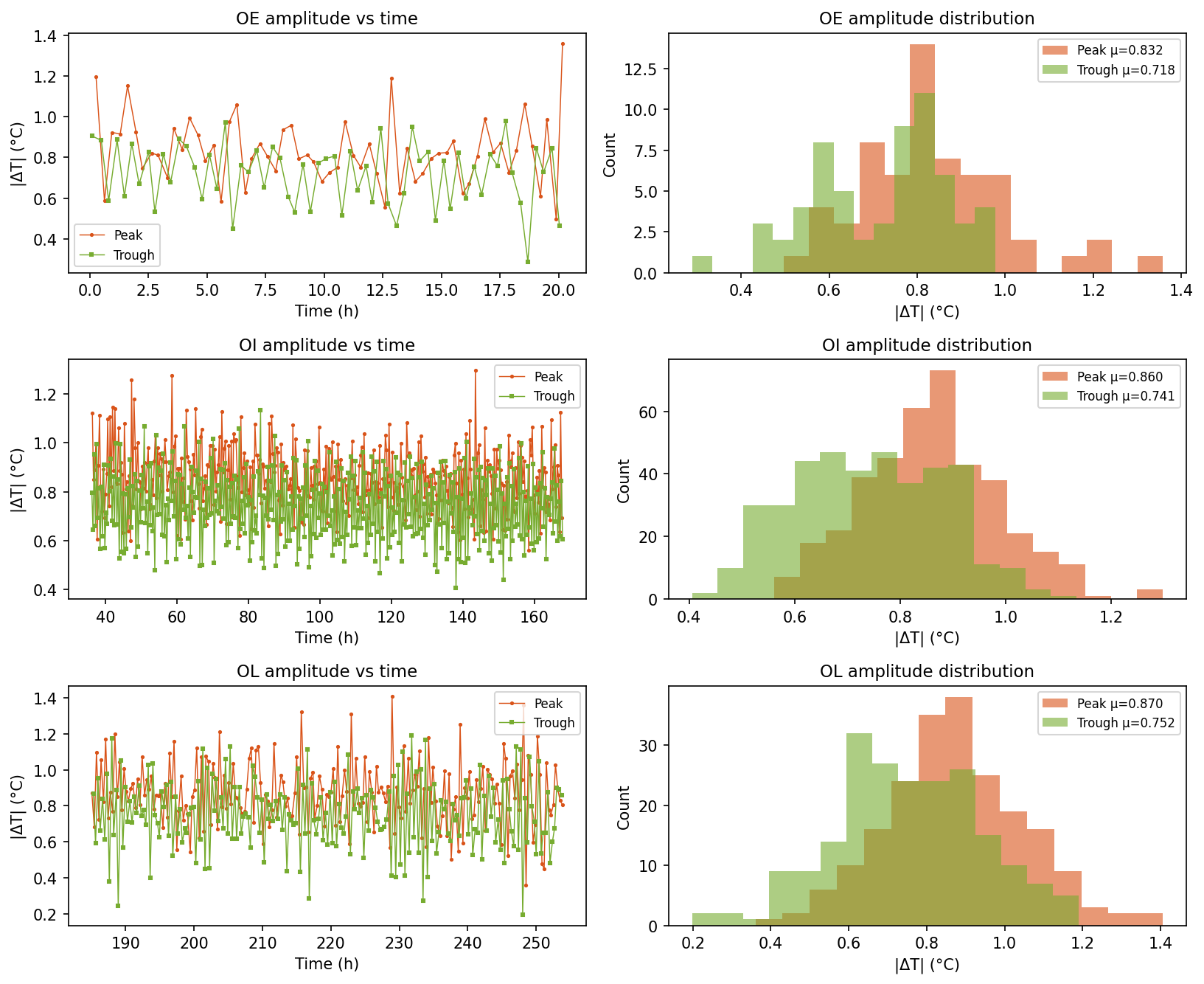}
  \caption{As Fig.~\ref{fig:ampA} for OE, OI, OL.}
  \label{fig:ampB}
\end{figure}

\begin{table}[htbp]
  \centering
  \caption{Amplitude statistics. CV: coefficient of variation.}
  \label{tab:amplitude}
  \begin{tabular}{l S[table-format=1.3] S[table-format=1.3]
                    S[table-format=1.3] S[table-format=1.3]
                    S[table-format=1.3]}
    \toprule
    & \multicolumn{2}{c}{Zoomed} & \multicolumn{3}{c}{Overview} \\
    \cmidrule(lr){2-3}\cmidrule(lr){4-6}
    Parameter & {ZE} & {ZL} & {OE} & {OI} & {OL} \\
    \midrule
    Peak $\mu$ (\si{\celsius})    & 0.983 & 1.043 & 0.832 & 0.860 & 0.870 \\
    Peak $\sigma$ (\si{\celsius}) & 0.140 & 0.184 & 0.162 & 0.127 & 0.176 \\
    Peak CV (\%)                  & 14.2  & 17.7  & 19.5  & 14.8  & 20.2  \\
    Trough $\mu$ (\si{\celsius})  & 0.753 & 0.806 & 0.668 & 0.701 & 0.680 \\
    Peak/Trough                   & 1.306 & 1.294 & 1.246 & 1.227 & 1.279 \\
    \bottomrule
  \end{tabular}
\end{table}

%% ─────────────────────────────────────────────────────────────────────────────
\section{Discussion}
\label{sec:discussion}

\subsection{Harmonic series}
\label{sec:harmonics}

A waveform with alternating half-periods $T_{\mathrm{charge}}$ and
$T_{\mathrm{discharge}}$ has fundamental $f_1 = (T_{\mathrm{charge}} +
T_{\mathrm{discharge}})^{-1}$ and generates harmonics $f_n = nf_1$ with
relative amplitudes encoding the waveform asymmetry~\cite{Bracewell2000}.
The dominant DFT peak at $f_2$ (\SI{\sim 20}{\minute}) is therefore the second
harmonic of the \SI{\sim 40}{\minute} fundamental, not an independent
oscillation, and the \SI{\sim 13}{\minute} feature is $f_3$.
Harmonic frequencies predicted from the interval analysis agree quantitatively
with the observed DFT peaks in all five segments.

\subsection{Relaxation pause vs.\ CCCV overhead}

The analysis of the fume hood temperature signal  supports strongly the hypothesis
that the signal corresponds to charge-recharge cycling at a 3C rate where
both CC charge and CC discharge nominally take \SI{20}{\minute}.
Two mechanisms could produce $T_{\mathrm{charge}} > T_{\mathrm{discharge}}$:
(i)~the CV tail of a CCCV charge protocol, and (ii)~an intentional relaxation
pause after charging.

Without access to the cycling protocol and further information on the structure of the devices, I
cannot favour either hypothesis over each other on the basis of the fume-hood signal alone.

\subsection{Degradation assessment}

The zoomed windows ZE and ZL, \SI{\approx 349} cycles apart within the same
run, show a \SI{0.96}{\minute} (\SI{2.4}{\percent}) increase in
$T_{\mathrm{cycle}}$ and a \SI{6.1}{\percent} increase in peak amplitude,
both statistically significant ($3.5\sigma$).
Their physical origin---ambient temperature drift, a protocol adjustment, or
minor impedance growth---cannot be inferred from the thermal record alone.
Within the overview segments (OE, OI, OL together covering \SI{332} cycles),
$T_{\mathrm{charge}}$, $T_{\mathrm{discharge}}$, and amplitude are all stable.
Overall, the cycling limit-cycle behaviour is well preserved across the full
\SI{254}{\hour} run, consistent with healthy device operation; quantitative
ageing projections are not warranted.

\subsection{Limitations and outlook}

The fume-hood air temperature is a filtered proxy for the device
surface temperature. Although the heat-generation rates and internal resistance cannot be
inferred without calibrating the thermal coupling the average temperature
difference between the contact sensor and the ambient sensor reflects the
coupling to some extent. The periodic signal is clearly resolved by a simple DFT analysis after
careful background subtraction, but is not the optimal tool for non-stationary or gapped signals of this
kind. More sophisticated approaches---Lomb–Scargle periodograms or wavelet
analysis--- could extract additional information.
Cross-correlation with simultaneously acquired electrical data would decouple
ohmic and entropic contributions and calibrate the fume-hood transfer function.
The use of the contact sensor demonstrated here for drift removal also opens
the possibility of differential measurements that further suppress common-mode
environmental noise.

%% ─────────────────────────────────────────────────────────────────────────────
\section{Conclusions}
\label{sec:conclusions}

Five segments of a single \SI{254}{\hour} fume-hood temperature record,
extracted from a published solid-state battery self-discharge
test~\cite{originalreport}, consistently reveal the thermal signature of the
co-located cycling devices of unknown chemistry:
$T_{\mathrm{charge}} \approx \SI{22}{\minute}$,
$T_{\mathrm{discharge}} \approx \SI{18}{\minute}$,
$T_{\mathrm{cycle}} \approx \SI{40}{\minute}$ (consistent with 3C),
and $\Delta \approx \SI{4}{\minute}$ (evidence for a fixed relaxation pause).
The Fourier spectrum is a harmonic series fully explained by this
alternating-period structure.
Amplitude and period are stable throughout the run with no detectable
degradation signature. The approach is entirely passive, requires only a commodity temperature sensor,
and leaves the device and protocol unmodified.

Direct manual counting of the peaks of the ambient temperature record in Figure 2 of VTT report~\cite{originalreport}
reveals that at least 338 charge/decharge test cycles were run on Donut Lab devices. Although the number of tested devices
cannot be determined from the temperature record alone the analysis suggest either 1) a consistent behaviour of a single device
or 2) very small device-to-device differences.  

%% ─────────────────────────────────────────────────────────────────────────────
\section*{Data availability}
The temperature data were extracted from the publicly available
report~\cite{originalreport}.
%Analysis scripts and extracted data arrays are available at
%[repository URL TBD].

\section*{Author contributions}
P.O.T.: conceptualisation, data extraction, formal analysis, writing.

\section*{Declaration of competing interest}
The author declares no competing interests.

\section*{Acknowledgements}
The author thanks VTT Technical Research Centre of Finland and Donut Lab for
permission to use the data from report~\cite{originalreport}. The author used Claude (Anthropic)
to assist in generating data analysis code and text drafting. 

%% ─────────────────────────────────────────────────────────────────────────────
\bibliographystyle{elsarticle-num}
\bibliography{references, refs2}

\end{document}